\documentstyle[12pt]{article}
\begin{document}
\title{Localization conditions for two-level systems}
\author{Hu Cheng, Mo-Lin Ge and Jiushu Shao \\
Nankai Insitute of Mathematics, Nankai University \\
Tianjin 300071, P. R. China}
\maketitle

\begin{abstract}
The dynamics of two-level systems in an external periodic field are
investigated in general. The necessary conditions of localization are
obtained through analysing the time-evolving matrix. It is found that
localization is possible if not only is the dynamics of the system periodic,
but also its period is the same as that of the external potential.
A model system in a periodic $\delta$-function potential is studied
thoroughly. \\


\end{abstract}
\newpage

\bibliographystyle{unsrt}
\newcommand{\bqna}{\begin{eqnarray}}
\newcommand{\eqna}{\end{eqnarray}}
\newcommand{\bqn}{\begin{equation}}
\newcommand{\eqn}{\end{equation}}
\newcommand{\crt}{\mbox{$c_{r}(t)$}}
\newcommand{\clt}{\mbox{$c_{l}(t)$}}
\newcommand{\dcr}{\mbox{${\dot{c}}_{r}(t)$}}
\newcommand{\dcl}{\mbox{${\dot{c}}_{l}(t)$}}
\newcommand{\bc}{\mbox{${\bf C}$}}
\newcommand{\bb}{\mbox{${\bf B}$}}
\newcommand{\dbc}{\mbox{$\dot{\bf C}$}}
\newcommand{\bm}{\mbox{${\bf M}$}}
\newcommand{\ba}{\mbox{${\bf A}$}}
\newcommand{\brl}{\mbox{${\bf R}_{l}$}}
\newcommand{\brr}{\mbox{${\bf R}_{r}$}}
\newcommand{\beg}{\mbox{${\bf e}_{\gamma}$}}
\newcommand{\bet}{\mbox{${\bf e}_{\theta}$}}
\newcommand{\gam}{\mbox{$\gamma_{min}$}}
\newcommand{\gams}{\mbox{$\gamma_{min}^{2}$}}
\newcommand{\dzo}{\mbox{$\Delta_{0}$}}
\newcommand{\dzos}{\mbox{$\Delta_{0}^{2}$}}
\newcommand{\omet}{\mbox{$\omega t$}}
\newcommand{\omtt}{\mbox{$\omega T$}}
\newcommand{\al}{\mbox{$\alpha$}}

Tunneling and localization play a central role in quantum
mechanics~\cite{b:qm}. They are the two aspects of one quantum phenomenon.
The two concepts are extremely important in solid state physics~\cite{a:al}.
The  great achievements of Anderson {\em et al.} are focused on the
localization of electrons while much progress on resonant tunneling has been
made since Esaki's group proposed the model of
superlattice~\cite{te:rt,cet:rtt}.
It is not until the discovery of the coherent destruction of
tunneling by H\"anggi's group~\cite{gjdh:t,gdjh:cd,gdh:st,gh:l}
that attention has been paid to the study of localization of a
single particle in the double-well potential under a periodic
acting field. Since this external control of tunneling will be practically
valuable in many research fields such as laser physics, chemical reaction,
macroscopic quantum mechanics
etc~\cite{cjmp:ic,h:psc,bm:li,xg:eo,doh:dt}, we need more knowledge about this
type of
localization or the suppression of tunneling. \\

To study the localization, the most popular model is a quartic double-well
system perturbed by a periodic monochromatic field.  Such a  model has been
investigated extensively using the Floquet theory~\cite{lb:qt,s:s}. Because
it is not analytically solvable, one often recourses to the numerical
calculation that
sometimes avoids physical insights. Recently it has been
demonstrated~\cite{gh:l,glp:tc,ws:ltl}
that a two-level system shows localization if the parameters of the
acting field are adjusted, which represents a common feature of the
double-well system.  In a recent paper~\cite{ws:ltl} it is shown that
the quantum dynamics of the
two-level system under a periodic external potential can be mapped to the
classical one of a charged particle moving in the harmonic oscillator
potential plus an magnetic field in a plane. The behavior of tunneling and
localization is fully described by the radial trajectory of the particle.
This gives us an interesting physical picture although the difficulties of
mathematical treatments are not lessened at all.\\

In this letter, we shall consider the dynamical behavior of two-level
systems in the external periodic field which is antisymmetric with
respect to time in one period. Although the dynamics has been generally
discussed through time-advancing matrix in~\cite{ws:ltl} there are still some
features unexplored. By the same procedure we will study the evolution of the
particle in the two states and give the necessary condition of localization.
We will also derive the explicit results when the external field is
a periodic $\delta$-function potential.\\

The Hamiltonian of the two-level system in the external periodic potential is
\bqn
\label{h:ex}
\hat{H}=-(\dzo/2)(|1><1|-|2><2|)+V(t)(|1><2|+|2><1|),
\eqn
where $\dzo$ is the energy splitting between the states $|1>$ and $|2>$,
$V(t)$ is the coupling between them induced by the external
periodic driving force, and \[V(T+t) = V(t), V(t) = -V(t+T/2),\]
with $T$ being the period of the external field. We call $V_(t)$ a
{\em generalized-parity} potential.\\

Define the left state and the right state as, respectively
\[
|l> \equiv (|1>+|2>)/\sqrt{2}, \]
and
\[
|r> \equiv (|1>-|2>)/\sqrt{2}. \]
The wave vector $|\Psi(t)>$ can be expanded in the basis $(|l>, |r>)$.
Denote
\bqn
\label{wf:lr}
|\Psi(t)> = c_{l}(t)|l>+c_{r}(t)|r>.
\eqn
Then $\bc \equiv (\clt, \crt)^{T}$ satisfy the equation of motion:
\bqn
\label{le:ds}
\dbc = \bm \bc,
\eqn
where
\[
\bm = \left( \begin{array}{cc}
              -i V(t) & i\dzo/2 \\
              i\dzo/2 & iV(t)
             \end{array} \right). \]
Since Tr{\bf M} = 0, the time-advancing mapping or propagator {\bf A} over a
single period is a 2D area-preserving one (det{\bf A}=1)~\cite{a:mm}.
Apparently this conclusion does not depend on the form of $V(t)$ .
Defining
\[ \ba(t): \bc(t)=\ba(t)\bc(0) \]
for $ 0\leq t < T$ and $\ba\equiv\ba(T)$, we have
\bqn
\label{te:eq}
 \bc(nT+t)=\ba(t)\ba^{n}\bc(0).
\eqn

Consider three states ${\bf C}(0)$, ${\bf C}(T/2)$ and ${\bf C}(T)$
at three times $0, T/2$ and $T$ respectively. The
initial state ${\bf C}(0)$
is arbitrary except that its components must satisfy the normalization
condition, $|c_{l}(0)|^{2}+|c_{l}(0)|^{2}=1$. We can write
\bqn\label{f:ma}
{\bf C}(T) = \left( \begin{array}{ll} \al & \beta \\
\gamma & \delta \\ \end{array} \right) {\bf C}(T/2),
\eqn
\bqn\label{s:ma}
{\bf C}(T/2) = \left( \begin{array}{ll} p & q \\
s & t \\ \end{array} \right) {\bf C}(0).
\eqn
According to the normalization condition, the matrix elements should satisfy
\bqn
|\al|^{2}+|\beta|^{2} = 1
\eqn
and
\bqn
|\gamma|^{2}+|\delta|^{2} = 1.
\eqn
Replacing $t$ by $-t$ in ~\ref{le:ds} and taking its complex conjugate we get
\bqn
\label{dbc:ss}
\dbc = \left( \begin{array}{cc}
              -i V(-t) & i\dzo/2 \\
              i\dzo/2 & iV(-t)
             \end{array} \right) {\bf C}.
\eqn
Note that $V(t)$ is antisymmetric with respect to $t=T/2$. If we move the
origin $t=0$ to $t=T/2$, then $V(t)=-V(t)$. Inserting it into~\ref{dbc:ss}
and after some analysis one can easily find the following relations
\[ \begin{array}{ll} p={\al}^{*}, & q=-{\gamma}^{*} \\
 s=-{\beta}^{*}, & t={\delta}^{*}.\\ \end{array} \]
Thus, we have
\bqn
\ba = \left(\begin{array}{cc} |\al|^{2}-|\beta|^{2} & {\beta}{\delta}^{*}
-{\al}{\gamma}^{*}\\ {\gamma}{\al}^{*}-{\delta}{\beta}^{*} & |\delta|^{2}
-|\gamma|^{2}\\ \end{array} \right).
\eqn
Together with~\ref{f:ma} and~\ref{s:ma}, we obtain
\[ \ba   =  \left( \begin{array}{cc}
                   a   &
                   -b+ic \\
                   b+ic &
                   a
                 \end{array}
          \right), \]
where $a, b$ and $c$ are real numbers determined by the system
and $a^{2}+b^{2}+c^{2}=1$ from det{\bf A}=1.

Because $-2\leq $ Tr$\ba=2a
\leq 2$, {\em the dynamics of the system is strongly stable}~\cite{a:mm}.
In other words, the future behavior of the two-level model is insensitive
to the initial condition. Roughly speaking, localization means that the state
of  the system evolves near the initial state. To investigate the rule of
localization, thus, we can choose any initial state as we like. But what
we are interested in practice is related to the evolution of either the left
state or the right one. Therefore, we suppose $\bc(0) =(1,0)^{T}$ in the
following discussion. The state at the $n$th stroboscopic
point $nT$ is given by
\bqn
\label{te:eq}
 \bc(nT)=\ba^{n}\bc(0).
\eqn
The power of $\ba$ can be determined by the \mbox{Caley-Hamilton}
theorem (see,  e.g.~\cite{vll:jap}):
\bqn
\label{ch:th}
\ba^{n-1}=P_{n-2}(a) \ba -P_{n-3}(a) {\bf I},
\eqn
where $P$ is the Chebyshev polynomial and ${\bf I}$ the identity matrix.
Denote $\sigma \equiv \arccos a$, then $P_{n}=\sin[(n+1)\sigma]/\sin\sigma$.\\
After a simple calculation we obtain the modulus of $c_{r}(nT)$:
\bqn
\left|c_{r}(nT)\right| = \left|\sin(n\sigma)\right| .
\eqn
Localization requires that for any integer $n$ the probability of finding the
system in
the right state should be small, i.e., $ |c_{r}(t)|<<1$. This is only a
qualitative statement. Quantitatively, we define localization as
$|c_{r}|<1/2 $ for all the time. As a necessary condition for localization,
therefore, we must have $|c_{r}(nT)|= |\sin(n\sigma)|<1/2$. Mathematically
if
\bqn
{\sigma}\neq\frac{Q\pi}{P}
\eqn
where $P$ and $Q$ are two mutually prime integers, then the set
$\{\sin(n\sigma): n{\in}N\}$ is dense in the interval [0, 1]. In other words,
 for
arbitrary $\sigma$ there must exist such an integer $n$ that $\sin(n\sigma)$
is very near to 1, as a consequence, localization is destroyed. To guarantee
localization, thus it is needed
\bqn
{\sigma}=\frac{Q\pi}{P}
\eqn
for some integers $P$ and $Q$, i.e., there must be a set of integers
$m$ to satisfy
\bqn\label{si:s}
\sin(m\sigma)=0.
\eqn
Suppose $m>1$ and $m$ is the smallest one in the set.
Thus $\sigma$ can be written as
\bqn
\sigma = \frac{l\pi}{m} ,
\eqn
where $l$ and $m$ are mutually prime integers. If $m$ is even, then
\bqn
\label{ls}
\left|\sin\left( \frac{m{\sigma}}{2} \right)\right| = 1,
\eqn
which is forbidden for localization because $mT/2$ is also a stroboscopic
point. If $m$ is odd, since $m$ and $l$ is mutually prime, there is an
integer $n$ less than $m$ such that  mod$(nl, m) = \frac{m-1}{2}$.
Then we have
\bqn
\label{lsi}
\left|\sin(n{\sigma})\right|^{2} =
\left|\sin\left[\frac{(m-1)\pi}{2}\right]\right|^{2}
> \frac{1}{2}.
\eqn
Taking account of~\ref{ls} and~\ref{lsi}, we conclude that only if $m =1$
localization may arise. Namely, we have proved that \\
{\em Theorem}: The necessary conditions for localization of two-level systems
in external periodic fields are 1) its dynamics should be periodic; 2) the
period should equal to that of the external fields. In mathematical language,
these conditions are simply described as
\bqn
\ba = \pm {\bf I}.
\eqn
This result strenghthens the conclusion in reference~\cite{ws:ltl} in which
only the first localization condition is given. Note that this observation
was obtained by Gro{\ss}mann and H\"anggi in~\cite{gh:l}, using the
Floquet approach. In the following, this
theorem will be applied to the treatment of the periodic $\delta$-function
potential or pulse acting on a two-level system. \\

The $\delta$-function potential is very important in physics because
it is simply representing the most physical features of other
acting potentials. We find that it is difficult to solve~\ref{le:ds} directly,
and so we treat the $\delta$-function field as the
limited case, $\epsilon \rightarrow 0$, of the rectangular
potential or pulse, namely,
\[ V(t) = \left\{ \begin{array}{ll}
                   0 & \mbox{if $0\leq t <T/4 -\epsilon$} \\
                   V_{0}/2\epsilon & \mbox{if $T/4-\epsilon\leq t
                   <T/4+\epsilon$} \\
                   0 & \mbox{if $T/4 +\epsilon \leq t <3T/4 -\epsilon $} \\
                   -V_{0}/2\epsilon & \mbox{if $3T/4 -\epsilon\leq t
                   <3T/4 +\epsilon$} \\
                    0  & \mbox{if $3T/4 +\epsilon\leq t <T$} \\
                    \end{array}
          \right., \]
and $V(T+t)=V(t)$ (see Fig. 1). If we
define the time-advancing matrix $\cal T$ of one $\delta$-function
potential $V_{0}{\delta}(t)$ as
\bqn
 \bc(0^{+})={\cal T}\bc(0^{-}),
\eqn
we can  derive
\bqn\label{aa:a}
{\cal T} = \left( \begin{array}{cc}
                   \exp(iV_{0})  & 0 \\
                   0 & \exp(-iV_{0})   \\
                   \end{array}
          \right).
\eqn
The time-advancing mapping {\bf A} turns out to be
\bqn
 \ba = \left( \begin{array}{cc}
                   1-2\sin^{2}\frac{T\dzo}{4}\cos^{2}V_{0}  &
                   -\sin2V_{0}\sin\frac{T\dzo}{4}+i\sin\frac{T\dzo}{2}
                   \cos^{2}V_{0} \\
                   \sin2V_{0}\sin\frac{T\dzo}{4}+i\sin\frac{T\dzo}{2}
                   \cos^{2}V_{0}  &
                   1-2\sin^{2}\frac{T\dzo}{4}\cos^{2}V_{0}  \\
                   \end{array}
          \right).
\eqn
{}From the definition of $\sigma$ we have
\bqn
\cos\sigma = 1-2\sin^{2}\frac{T\dzo}{4}\cos^{2}V_{0}.
\eqn
Using the theorem above, we find the necessary condition for localization as
\bqn
\sin\frac{T\dzo}{4}\cos{V_{0}} = 0,
\eqn
or
\bqn
\left\{
\begin{array}{l} \sin{V_{0}} = 0 ,\\
\cos\frac{T\dzo}{4} = 0.\\ \end{array} \right.
\eqn
If this condition is satisfied, and if the system is
localized in one single period, then localization will dominate the dynamics
for the whole time. In order to find the sufficient condition for
localization, therefore, we only need study the dynamical features over one
period. To this end, it is required that
the modulus of $c_{l}$ be never less than a half in one period. This
requirement imposes constraints on $V_{0}$ and $T$ respectively, i.e.,
\bqn
\label{qqq}
\left\{ \begin{array}{ll}V_{0} = (n+\frac{1}{2})\pi ,\\
0 \leq T\dzo \leq 2\pi ,\\ \end{array} \right.
\eqn
where $n$ is an integer. This gives the sufficient conditions of
localization  for a two-level system in the periodic $\delta$-function
field. \\

Returning to~\ref{aa:a}, we see clearly that the applying $\delta$-function
pulse only brings about changes in
the phases of the left and the right states. The former is increased by
$V_{0}$ and the latter decreased by $V_{0}$. Since
the evolution is strongly dependent on the difference of the two phases,
we can always choose a series of appropriate $\delta$-function potentials to
obtain localization. For example, let us apply a periodic
potential consisting of a pair of $\delta$-function pulses with opposite
signs in one period to a two-level system. The two
pulses are assumed to take effect at one fourth period and three fourths
period (see Fig. 2). Suppose the system is in the left state at the
beginning. As shown above, if and only if the positive and the negative
pulses have the same amplitudes, $|V_{0}|=(n+\frac{1}{2})\pi$ and
the period $T$ is not larger than ${2\pi}/{\dzo}$, then the system
is localized and the probability for the system to stay in the left state
will not be less than $\rho_{l, min}=\cos^{2}({T\dzo}/{8})$. Noticing that
$T\leq {2\pi}/{\dzo}$, we find that $\rho_{l, min}$ is a monotonically
decreasing function of $T$, which indicates that the smaller $T$ is, the
larger $\rho_{l, min}$. In other words, we can realize strong localization
by taking short-period $\delta$-function pulses. The varying of the
probability according to time for different periods is illustrated in
Fig. 3. It should be stressed that although the period of the dynamics
is identical with that of the external field, the period of the evolution
of the probability in the left state is a half.
In fact, this is a common feature of the two-level system perturbed by
any external potential possessing the generalized parity, which can be
shown via symmetry analysis~\cite{ws:ltl}. \\


In conclusion, we have derived the necessary conditions for a two-level
system driven by a periodic field to be localized: not only must the
system evolve periodically, but also the period is identical with
that of the external potential. The necessary conditions can always
be satisfied by adjusting the period and the strength of the applying field.
Under these conditions the system will get localized if and only if it is
localized over one single period, which becomes possible through
modifying the parameters of the external field. In the case of
$\delta$-function pulses, the necessary and the sufficient conditions
for localization are acquired analytically. In principle,
any applying potential can be imitated approximately in terms of
a series of $\delta$-function pulses, and so the general localization
conditions can be investigated on the basis of the results from our
$\delta$-function model. Besides, we expect that this
simple theory can be employed to study the process of enantiomerization,
which is of much significance in understanding the origin of life.\\

\newpage

\noindent Captions of figures\\

\noindent Fig. 1 \hspace*{1.0cm} An applied field
consisting of rectangular pulses. It becomes
the $\delta$-function field by taking the limit
$\epsilon \rightarrow 0$.\\

\noindent Fig. 2 \hspace*{1.0cm} Schematic graph of applied $\delta$-function
pulses.\\

\noindent Fig. 3 \hspace*{1.0cm}Calculated probabilities in the left state
versus time
for three periods. Here the solid line corresponds to $T\dzo = 4\pi/5$, the
dot dashed line $T\dzo = 4\pi/3$ and the dashed line $T\dzo = 2\pi$.\\

\newpage
 \hspace*{4cm} $V(t)$ \hspace*{4cm} $V(t)$ \vspace*{3cm}\\
 \hspace*{4cm} $V(t)$ \hspace*{4cm} $V(t)$ \vspace*{3cm}\\
time (unit: 2/{\dzo}) \hspace*{1cm} 2T \hspace*{1cm} T/2 \hspace*{1cm}
{\bf 0} \vspace*{2.0cm}\\
T/4 \hspace*{1cm} 3T/4 \hspace*{1cm} 2T \hspace*{1cm} {\bf 0} \hspace*{1cm}
$2{\bf \epsilon}$ \hspace*{1cm} $V_{0}/{2\epsilon}$ \vspace*{2cm}\\

\large probability \vspace*{2.0cm} in the left state \hspace*{2cm} $V_(t)$
\hspace*{2cm} $-V_{0}$\\

time (unit: 2/{\dzo}) \hspace*{1cm} T \hspace*{1cm} T/2 \hspace*{1cm}
{\bf 0} \vspace*{2.0cm}\\
T/4 \hspace*{1cm} 3T/4 \hspace*{1cm} T \hspace*{1cm} {\bf 0} \hspace*{1cm}
$2{\bf \epsilon}$ \hspace*{1cm} $V_{0}/{2\epsilon}$\\

\end{document}